\documentclass[aps,prb,twocolumn,showpacs,amsfonts]{revtex4}

\usepackage{epsfig}
\usepackage{amsmath}
\usepackage{amssymb}
\usepackage{bm}
\usepackage{color}

\newcommand{\beq}{\begin{equation}}
\newcommand{\eeq}{\end{equation}}
\newcommand{\beqarray}{\begin{eqnarray}}
\newcommand{\eeqarray}{\end{eqnarray}}

\newcommand{\Hc}{\ensuremath{\mbox{H.c.}}} 
\newcommand{\eq}[1]{Eq.~(\ref{#1})} 
\newcommand{\fig}[1]{Fig.~\ref{#1}} 
\newcommand{\Ref}[1]{Ref.~\onlinecite{#1}} 

\begin{document}

\allowdisplaybreaks

\title{Microscopically derived Ginzburg-Landau theory for magnetic order
  in the iron pnictides}

\author{P. M. R. Brydon}
\email{brydon@theory.phy.tu-dresden.de}
\author{Jacob Schmiedt}
\author{Carsten Timm}
\email{carsten.timm@tu-dresden.de}
\affiliation{Institut f\"{u}r Theoretische Physik, Technische Universit\"{a}t  
  Dresden, 01062 Dresden, Germany }

\date{November 15, 2011}

\begin{abstract}
We examine the competition of the observed stripe spin density wave (SDW) with
other commensurate and incommensurate SDW phases in a two-band model of the
pnictides. Starting from this microscopic model, we rigorously derive an
expansion of the free energy in terms of the different order parameters at the
mean-field level. We show that three distinct
commensurate SDW states are possible and study their appearance as a
function of the doping and the electronic structure. We show that the stripe
phase is generally present, but its extent in the phase diagram depends
strongly upon the number of hole Fermi pockets that are nested with the
electron Fermi pockets. Electron pockets competing for the same
portion of a hole pocket play a crucial role. We discuss the
relevance of our results for the antiferromagnetism of the pnictides.
\end{abstract}


\maketitle

\section{Introduction}

The superconductivity of the iron pnictides continues to fascinate
the condensed matter community.\cite{Paglione2010,Johnston2010,Lumsden2010}
Because of their high critical temperatures, particular 
interest has focused upon the so-called 1111 family\cite{Kamihara2008} 
{\it R}FeAsO and the 122 family\cite{Rotter2008} $A{\rm Fe_2As_2}$ ({\it R} and
{\it A} are rare-earth and alkaline-earth elements, respectively), which
become superconducting by chemical doping or under pressure. The parent
compounds are antiferromagnets,~\cite{Lumsden2010,Johnston2010} with
stripe-like magnetic order with respect to the 
lattice of Fe atoms. Furthermore, the antiferromagnetism is intimately linked
to an 
orthorhombic distortion of the crystal,~\cite{1111coupling,122coupling}
as evidenced by the coincidence of the ferromagnetic direction with 
the shorter crystallographic axis in all 1111 and 122 parent
compounds. It has been argued that the same condition that favors
stable stripe order
also implies a nematic transition above the N\'{e}el temperature $T_N$,
where the
magnetic fluctuations on each sublattice become locked into a stripe
configuration,\cite{Fang2008,Kim2011,Fernandes2010} and which produces the
orthorhombic distortion via the 
magnetoelastic coupling.~\cite{Kim2011,BG2009,magph} The mechanism for
stabilizing the stripe order is therefore a key 
problem in pnictide physics. 

The microscopic origin of the antiferromagnetism in the pnictides has been
approached in a number of different ways. Frustrated local moment models for
the Fe spins can reproduce the observed magnetic
order,~\cite{Johnston2010,localspin,localspinorb} but the evidence for the
metalicity~\cite{McGuire,Liu2008,Dong2008} and relatively weak
correlations~\cite{WLYang2009} of the  
parent compounds, and the development of incommensurate (IC) magnetic order
upon doping,~\cite{Pratt2011} suggest an itinerant description. \emph{Ab
  initio} calculations predict,~\cite{nesting,Zhang2010} and angle-resolved 
photoemission and magneto-oscillation 
experiments confirm,~\cite{magneto,ARPES} that the Fermi 
surface of the pnictide parent compounds have quasi-two-dimensional nested
electron and hole pockets. Such a system is known to undergo an excitonic
instability towards a spin-density-wave (SDW)
state,~\cite{Excitonic,Buker1981} as for example in
chromium.~\cite{Rice1970,Fawcett1988} Many 
authors have in addition emphasized the importance of the complicated orbital
structure of the Fermi
surfaces,~\cite{Kuroki2008,Graser2009,Raghu2008,Nicholson2011,Lorenzana2008,Raghuvanishi2011,KEM2011,Brydon2011,Schmiedt2011} 
but key aspects of the physics are nevertheless well
understood on the basis of simpler orbitally trivial ``excitonic'' 
models.~\cite{Brydon2009b,KECM2010,EC2010,KEM2011,FS2010,VVC,ZTB2011,Chubukov2008,CT2009,Brydon2009a,QI,KESM2011,KT2011,PHW2009,MC2010} 

Most itinerant models of the pnictides display at least two nesting
instabilities at different wavevectors. There is hence competition between
the stripe magnetic order and other  SDW
phases. Within a minimal two-orbital model,~\cite{Raghu2008} it has been shown
that doping away from half-filling~\cite{Lorenzana2008,Raghuvanishi2011} or
a relatively large ratio of the Hund's rule coupling to the Coulomb
repulsion~\cite{Raghuvanishi2011} can stabilize 
the stripe state. For excitonic models, on the other hand, Eremin and
Chubukov~\cite{EC2010} 
have demonstrated that the ellipticity of the electron pockets or interactions 
between the electron bands can stabilize the observed SDW state.
The stripe order was nevertheless found to be rather 
sensitive to the number of Fermi pockets involved in the SDW, and its extent in
the phase diagram remains uncertain. Competition with a different excitonic
instability has also been proposed to stabilize a stripe SDW.~\cite{KT2011}

In this paper we present a systematic study of the magnetic order in the
popular two-band excitonic model of the 
pnictides.~\cite{Brydon2009b,KEM2011,Nicholson2011,KESM2011,FS2010,VVC} 
Keeping only interaction terms which lead to an excitonic state, 
we construct the Dyson equation for the single-particle Green's
functions in an arbitrary commensurate SDW 
phase treated at the mean-field level.
By iterating the Dyson equation, we obtain approximate forms for the
self-consistency equations for the order parameters valid near $T_N$, which we
then integrate to obtain a Ginzburg-Landau expansion of the free energy. From
this we determine the phase diagram for several choices of the
non-interacting band structure, and show that three different commensurate SDW
states are possible. We conclude with a discussion of the relevance
to the magnetism of the pnictide parent compounds.

\section{Model}

\begin{figure}
\begin{center}
\includegraphics[clip,width=\columnwidth]{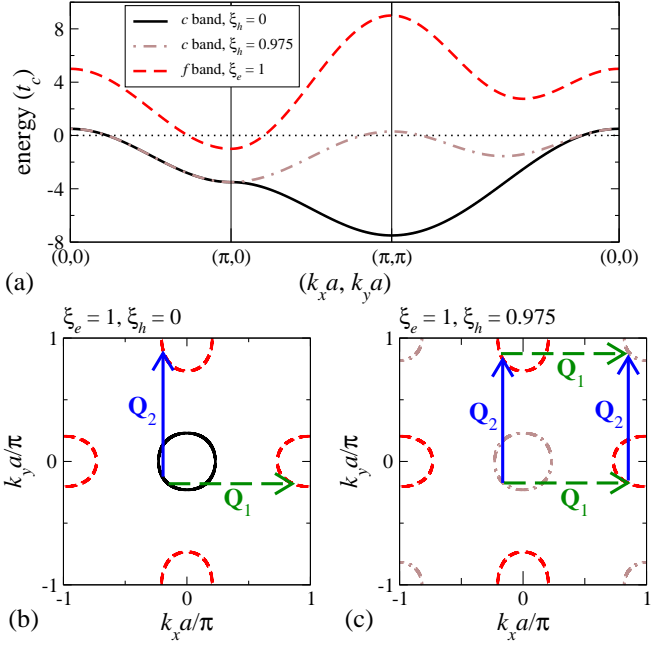}
\end{center}
\caption{\label{bs}(color online) (a) Dispersion of the electron and hole
  bands for the cases of a single hole pocket at the $\Gamma$ point ($\xi_h=0$)
  and of hole pockets at both the $\Gamma$ and M points ($\xi_h=0.975$). Panels
  (b) and (c) show the corresponding Fermi surfaces and
  the nesting vectors ${\bf Q}_1=(\pi/a,0)$ and ${\bf
  Q}_2=(0,\pi/a)$. In all panels we set $\mu=0$.} 
\end{figure}

We model the FeAs planes as a two-dimensional interacting
two-band system, where one band has electron-like Fermi pockets, while the
other has hole-like Fermi pockets. Including only interaction terms which
directly lead to an excitonic instability, we write the Hamiltonian as
\beqarray
H &=&\sum_{{\bf{k}},\sigma}\left\{(\epsilon^{c}_{{\bf{k}}}-\mu)c^{\dagger}_{{\bf{k}}\sigma}c^{}_{{\bf{k}}\sigma}
+
(\epsilon^{f}_{{\bf{k}}}-\mu)f^{\dagger}_{{\bf{k}}\sigma}f^{}_{{\bf{k}}\sigma}
\right\} \notag \\
& & {}+
\frac{g_{1}}{V}\sum_{{\bf{k}},{\bf{k}}',{\bf{q}}}\sum_{\sigma,\sigma'}
c^{\dagger}_{{\bf{k}}+{\bf{q}},\sigma}c^{}_{{\bf{k}}\sigma}f^{\dagger}_{{\bf{k}}'-{\bf{q}},\sigma'}f^{}_{{\bf{k}}'\sigma'}
\notag \\
&& {}+
\frac{g_{2}}{V}\sum_{{\bf{k}},{\bf{k}}',{\bf{q}}}\left\{c^{\dagger}_{{\bf{k}}+{\bf{q}},\uparrow}c^{\dagger}_{{\bf{k}}'-{\bf{q}},\downarrow}f^{}_{{\bf{k}}',\downarrow}f^{}_{{\bf{k}},\uparrow}
+ \Hc
  \right\} \, , \label{eq:Ham} 
\eeqarray
where $c^{\dagger}_{{\bf{k}}\sigma}$ ($f^{\dagger}_{{\bf{k}}\sigma}$)
creates a spin-$\sigma$ electron with momentum ${\bf{k}}$ in the hole-like
(electron-like) band. In terms of the single-Fe unit cell, we assume the
dispersions   
$\epsilon^{c}_{{\bf{k}}} = \epsilon_{c} + 2t_{c}(1-\xi_h)[\cos(k_{x}a) +
  \cos(k_{y}a)] + 2t_c\xi_h[1 + \cos(k_xa)\cos(k_ya)]$ and
$\epsilon^{f}_{{\bf{k}}} = \epsilon_{f} + t_{f,1}\cos(k_{x}a)\cos(k_{y}a) -
t_{f,2}\xi_e[\cos(k_{x}a)+\cos(k_{y}a)]$, where $a$ is the Fe-Fe bond
length. In units of $t_c$, we take $\epsilon_{c}=-3.5$, $\epsilon_{f}=3.0$,
$t_{f,1}=4.0$, and $t_{f,2}=1.0$. We plot representative band structures and
Fermi surfaces 
in~\fig{bs}. The dimensionless 
quantities $\xi_{e}$ and $\xi_h$ are key control parameters:
$\xi_e$ controls the ellipticity of the electron pockets, while varying
$\xi_{h}$ from $0$ to $1$ tunes the band structure from a system with a
single hole pocket at the $\Gamma$ point to a system with equally large hole
pockets at both the $\Gamma$ and the M points. For $\xi_h\approx0$ each electron 
pocket is nested with the hole pocket by only one of the orthogonal
wavevectors ${\bf Q}_{1}=(\pi/a,0)$ and ${\bf Q}_{2}=(0,\pi/a)$ [\fig{bs}(b)],
while for 
$\xi_h\approx1$ both electron pockets can nest with a hole pocket at each
nesting vector [\fig{bs}(c)]. A system with a single hole pocket and two
electron pockets was proposed as a minimal model of the pnictides
in~\Ref{EC2010}, and has been examined by a number of 
authors.~\cite{Brydon2009b,KECM2010,KESM2011,MC2010} On the other hand, a
Fermi surface with  
hole pockets at the $\Gamma$ and M points is realized in the minimal
two-\emph{orbital}
model of the pnictides~\cite{Raghu2008} and this situation has been extensively
studied.~\cite{Raghu2008,Lorenzana2008,Nicholson2011,KEM2011,MC2010,EC2010,Raghuvanishi2011}
Furthermore, 
it is also of relevance to more sophisticated orbital models where in addition
to the $d_{xz}/d_{yz}$-derived hole pockets at the $\Gamma$ point there is
usually also a $d_{xy}$-derived hole pocket at the M
point,~\cite{Graser2009,Kuroki2008} which may play an 
important role in generating the SDW order.~\cite{Brydon2011,Schmiedt2011}
Although the orbital content of the $\Gamma$ and M hole pockets are very
different, mean-field studies suggest that the SDW instability is primarily
determined by the nesting properties,~\cite{Brydon2011,Schmiedt2011} hence
justifying the orbitally-trivial excitonic model used here. 

Equation (\ref{eq:Ham}) contains a density-density interaction
and a term describing correlated transitions between the electron and
hole bands, with contact potentials $g_{1}$ and $g_{2}$,
respectively. At sufficiently low temperatures, the system is unstable against
an excitonic SDW with effective coupling
$g_{\text{SDW}}=g_{1}+g_{2}>0$.~\cite{Chubukov2008,Buker1981,Brydon2009b}  
For our system the excitonic SDW
state has two order parameters corresponding to electron-hole pairing with
relative wavevector equal to ${\bf Q}_1$ and ${\bf Q}_2$, i.e., ${\bf
  \Delta}_{1} = \sum_{\alpha,\beta}{\bf \Delta}_{1,\alpha,\beta} =
(1/V)\sum_{\bf{k}}\sum_{\alpha,\beta}\hat{\pmb 
  \sigma}_{\alpha,\beta}\langle{c^{\dagger}_{{\bf{k}}+{\bf{Q}}_{1},\alpha}f^{}_{{\bf{k}}\beta}}\rangle$
and ${\bf \Delta}_{2} = \sum_{\alpha,\beta}{\bf \Delta}_{2,\alpha,\beta}
  = (1/V)\sum_{\bf{k}}\sum_{\alpha,\beta}\hat{\pmb
  \sigma}_{\alpha,\beta}\langle{c^{\dagger}_{{\bf{k}}+{\bf{Q}}_{2},\alpha}f^{}_{{\bf{k}}\beta}}\rangle$,
where $\hat{\pmb \sigma}$ is the vector of the Pauli matrices. 
${\bf \Delta}_{1}$ and ${\bf \Delta}_{2}$ are related to
the magnetization of each Fe sublattice by
$\mathbf{m}_a = \mathbf{\Delta}_1 + \mathbf{\Delta}_2$,
$\mathbf{m}_b = \mathbf{\Delta}_1 - \mathbf{\Delta}_2$.
When both ${\bf \Delta}_{1}$ and ${\bf \Delta}_{2}$ are non-zero, therefore,
the magnetization is the superposition of two SDW states with
orthogonal ordering vectors. It has been pointed out that
in the case that ${\bf \Delta}_{1}\cdot{\bf
  \Delta}_{2}\neq0$ one has to introduce additional charge-density-wave (CDW)
order parameters $\delta_{c}=(1/V)\sum_{{\bf
    k},\sigma}\langle{c^{\dagger}_{{\bf{k}}+{\bf{Q}}_{3},\sigma}c^{}_{{\bf{k}}\sigma}}\rangle$
and $\delta_{f}=(1/V)\sum_{{\bf
    k},\sigma}\langle{f^{\dagger}_{{\bf{k}}+{\bf{Q}}_{3},\sigma}f^{}_{{\bf{k}}\sigma}}\rangle$,
where ${\bf Q}_{3}={\bf Q}_1+{\bf Q}_2$.~\cite{Lorenzana2008,BBZ2010}

\section{Free energy expansion}

We define the single-particle Green's functions of the excitonic SDW system by
\beq
G^{a,b}_{{\bf Q},\sigma,\sigma'}({\bf k},i\omega_n) = -\int^{\beta}_{0}d\tau
\langle{T_{\tau} a^{}_{{\bf k}+{\bf Q},\sigma}(\tau)b^{\dagger}_{{\bf
      k},\sigma'}(0)}\rangle e^{i\omega_n\tau} \, ,
\eeq
where $a,b=c,f$. Treating the SDW and CDW orders at the mean-field level, we write
the Dyson equation for the Green's functions as
\begin{widetext}
\beqarray
G^{a,b}_{{\bf Q}_{n},\sigma,\sigma'}({\bf k},i\omega_n) & = &
\delta_{a,b}\delta_{\sigma,\sigma'}\delta_{{\bf Q}_n,0}G^{a (0)}({\bf k},i\omega_n) +g_{1}\delta_{\overline{a}}G^{a (0)}({\bf k}+{\bf Q}_{n},i\omega_n)G^{a,b}_{{\bf Q}_{n}+{\bf
    Q}_3,\sigma,\sigma'}({\bf k},i\omega_n) \notag \\
&& -g_{\text{SDW}}\sum_{m=1,2}\sum_{\alpha}{\bf \Delta}_{m,\alpha,\sigma}G^{a (0)}({\bf k}+{\bf Q}_{n},i\omega_n)G^{\overline{a},b}_{{\bf Q}_{n}+{\bf
    Q}_m,\alpha,\sigma'}({\bf k},i\omega_n) \, ,\label{eq:Dyson}
\eeqarray
\end{widetext}
where $\overline{c}=f$ and $\overline{f}=c$, and the
Green's functions of the 
non-interacting system are $G^{a (0)}({\bf k},i\omega_n)=(i\omega_n -
  \epsilon^{a}_{\bf k} + \mu)^{-1}$. The order
parameters can be expressed in terms of the Green's functions as
\begin{subequations} \label{eq:gapequations}
\beqarray
{\bf \Delta}_{m,\sigma,\sigma'} & = & \frac{1}{V}\sum_{{\bf
  k}}\frac{1}{\beta}\sum_{n}{\cal{G}}^{f,c}_{{\bf Q}_m,\sigma',\sigma}({\bf
  k},i\omega_n)e^{i\omega_n0^+} , \qquad \\
\delta_{\nu=c,f} & = & \frac{1}{V}\sum_{{\bf
  k},\sigma}\frac{1}{\beta}\sum_{n}{\cal{G}}^{\nu,\nu}_{{\bf
    Q}_3,\sigma,\sigma}({\bf k},i\omega_n)e^{i\omega_n0^+} .
\eeqarray
\end{subequations}
where $\beta=1/k_BT$.
In general, it is not possible to analytically solve~\eq{eq:Dyson} for the
Green's functions. By iterating the Dyson equation, however, we are able to
expand the Green's functions in terms of the order parameters. Inserting this
expansion into the self-consistency (``gap'') equations~(\ref{eq:gapequations}),
and truncating it above a given order, we 
hence obtain an approximate form of the self-consistency equations valid close to $T_N$
(assuming a second-order transition, as is the case here).
Since the self-consistency equations are obtained from the stationary points of the free
energy with respect to the order parameters, we can construct a Ginzburg-Landau
expansion for the free energy $F$ by integrating them,
\beqarray
F &=& F_{0} + \alpha\left(|{\bf \Delta}_{1}|^2 + |{\bf \Delta}_{2}|^2\right) +
\beta_{0}\left(|{\bf \Delta}_{1}|^4 + |{\bf \Delta}_{2}|^4\right) \notag \\
&&+ \beta_{1}|{\bf \Delta}_{1}|^2|{\bf \Delta}_{2}|^2 +
\beta_{2}|{\bf \Delta}_{1}\cdot{\bf \Delta}_{2}|^2 \notag \\
&& + (\gamma_{c}\delta_c + \gamma_{f}\delta_f){\bf \Delta}_{1}\cdot{\bf
  \Delta}_{2} + \alpha_{cf}\delta_{c}\delta_f +
\alpha_{c}\delta_c^2 +
\alpha_{f}\delta_f^2 \, , \notag\\\label{eq:freeE} 
\eeqarray
where $F_{0}$ is independent of the order parameters. We keep only
second-order terms involving the CDW order parameters, since the system is
far away from a pure CDW instability and a CDW only emerges as a secondary
order parameter.\cite{Toledano} We also neglect gradient
terms since we are only interested in homogeneous 
states. The coefficients in~\eq{eq:freeE} are written in terms of the
non-interacting Green's functions as follows:
\begin{widetext}
\begin{subequations} \label{eq:coefficients}
\beqarray
\alpha & = & 2g_{\text{SDW}}\left[1 + \frac{g_{\text{SDW}}}{V}\sum_{\bf
  k}\frac{1}{\beta}\sum_{n}G^{c (0)}({\bf k},i\omega_n)G^{f (0)}({\bf
  k}+{\bf Q}_1,i\omega_n)\right] \, , \\
\beta_0 & = & \frac{g_{\text{SDW}}^4}{V}\sum_{\bf
  k}\frac{1}{\beta}\sum_{n}\left[G^{c (0)}({\bf k},i\omega_n)G^{f (0)}({\bf
  k}+{\bf Q}_1,i\omega_n)\right]^2 \, , \label{eq:beta} \\
\beta_{1} & = & \frac{2g_{\text{SDW}}^4}{V}\sum_{\bf
  k}\frac{1}{\beta}\sum_{n}\left\{G^{c (0)}({\bf k},i\omega_n)G^{c (0)}({\bf
  k}+{\bf Q}_3,i\omega_n)\left[G^{f (0)}({\bf
  k}+{\bf Q}_1,i\omega_n)\right]^2\right. \notag \\
&& {}+ G^{f (0)}({\bf k}+{\bf Q}_{1},i\omega_n)G^{f (0)}({\bf
  k}+{\bf Q}_2,i\omega_n)\left[G^{c (0)}({\bf
  k},i\omega_n)\right]^2 \notag \\
&& \left. {}-  G^{c (0)}({\bf k},i\omega_n)G^{c (0)}({\bf
  k}+{\bf Q}_3,i\omega_n)G^{f (0)}({\bf
  k}+{\bf Q}_1,i\omega_n)G^{f (0)}({\bf
  k}+{\bf Q}_2,i\omega_n)\right\} \, , \label{eq:beta1} \\
\beta_{2} & = & \frac{4g_{\text{SDW}}^4}{V}\sum_{\bf
  k}\frac{1}{\beta}\sum_{n}G^{c (0)}({\bf k},i\omega_n)G^{c (0)}({\bf
  k}+{\bf Q}_3,i\omega_n)G^{f (0)}({\bf
  k}+{\bf Q}_1,i\omega_n)G^{f (0)}({\bf
  k}+{\bf Q}_2,i\omega_n) \, ,\\
\gamma_{c} & = & \frac{4g_{1}g_{\text{SDW}}^2}{V}\sum_{\bf
  k}\frac{1}{\beta}\sum_{n}G^{c (0)}({\bf k},i\omega_n)G^{f (0)}({\bf
  k}+{\bf Q}_1,i\omega_n)G^{f (0)}({\bf k}+{\bf Q}_2,i\omega_n) \, , \\ 
\gamma_{f} & = & \frac{4g_{1}g_{\text{SDW}}^2}{V}\sum_{\bf
  k}\frac{1}{\beta}\sum_{n}G^{c (0)}({\bf k},i\omega_n)G^{c (0)}({\bf
  k}+{\bf Q}_3,i\omega_n)G^{f (0)}({\bf
  k}+{\bf Q}_1,i\omega_n) \, , \\
\alpha_{cf} & = & g_{1} \, , \\
\alpha_{c} & = & -\frac{g_{1}^2}{V}\sum_{\bf
  k}\frac{1}{\beta}\sum_{n}G^{f (0)}({\bf k}+{\bf Q}_1,i\omega_n)G^{f (0)}({\bf
  k}+{\bf Q}_2,i\omega_n) \, , \\
\alpha_{f} & = & -\frac{g^2_{1}}{V}\sum_{\bf
  k}\frac{1}{\beta}\sum_{n}G^{c (0)}({\bf k},i\omega_n)G^{c (0)}({\bf
  k}+{\bf Q}_3,i\omega_n) \, .
\eeqarray
\end{subequations}
\end{widetext}
The CDW order parameters can be integrated out, resulting in the
renormalization of
\beq
\beta_{2} \rightarrow \widetilde{\beta}_{2} = \beta_{2} +
\frac{\alpha_{c}\gamma_f^2 + \alpha_f\gamma_c^2 - \alpha_{cf}\gamma_c\gamma_f}{\alpha_{cf}^2 -
  4\alpha_c\alpha_f} \, .
\eeq
The Ginzburg-Landau expansion of the free energy, \eq{eq:freeE}, and the
expressions for the coefficients in terms of a specific microscopic
model, \eq{eq:coefficients}, are the first major results of our paper.

\section{Phase Diagram}

The free energy in~\eq{eq:freeE} admits three possible commensurate SDW states
which we name following~\Ref{Lorenzana2008}:
\begin{itemize}
\item A magnetic stripe (MS) state where only one of the excitonic order
  parameters is nonzero, e.g., ${\bf \Delta}_{1}\neq0$, ${\bf
    \Delta}_{2}=0$. This corresponds to the ordering in the pnictides. This
  state minimizes the free energy if
  $2\beta_0<\min\{\beta_{1}+\widetilde{\beta}_{2},\beta_{1}\}$. 
\item An orthomagnetic (OM) state where $|{\bf \Delta}_{1}|=|{\bf \Delta}_2|$
  and 
  ${\bf \Delta}_1\perp {\bf \Delta}_2$. This corresponds to a ``flux'' type
  ordering of the magnetic moments. This
  state minimizes the free energy if
  $\beta_{1}<\min\{2\beta_0,\beta_{1}+\widetilde{\beta}_{2}\}$. 
\item A spin and charge order (SCO) state where $|{\bf \Delta}_{1}|=|{\bf
  \Delta}_2|$ and 
  ${\bf \Delta}_1\| \pm {\bf \Delta}_2$. In this state only
  one sublattice of the Fe plane has non-zero moments, which order in a
  checkerboard pattern. When 
  $g_{1}\neq0$, the spin order \emph{induces} a CDW with ordering vector ${\bf 
    Q}_3$. This
  state minimizes the free energy if
  $\beta_{1}+\widetilde{\beta}_{2}<\min\{2\beta_0,\beta_{1}\}$. 
\end{itemize}
From close examination of Eqs.\ (\ref{eq:beta}) and (\ref{eq:beta1}) we
observe that if 
$\epsilon^{f}_{\bf k}=\epsilon^{f}_{{\bf 
    k}+{\bf Q}_3}$ or $\epsilon^{c}_{\bf k}=\epsilon^{c}_{{\bf 
    k}+{\bf Q}_3}$ for \emph{all} ${\bf k}$ we have $2\beta_{0}=\beta_{1}$,
and hence the MS and OM states are degenerate. These conditions are satisfied
for the electron and hole bands at $\xi_e=0$ and $\xi_{h}=1$,
respectively. In particular, if $\xi_h\neq1$ the
degeneracy of the MS and OM states is lifted by arbitrarily small
ellipticity of the electron Fermi pockets, as pointed out in~\Ref{EC2010}. 

\begin{figure}
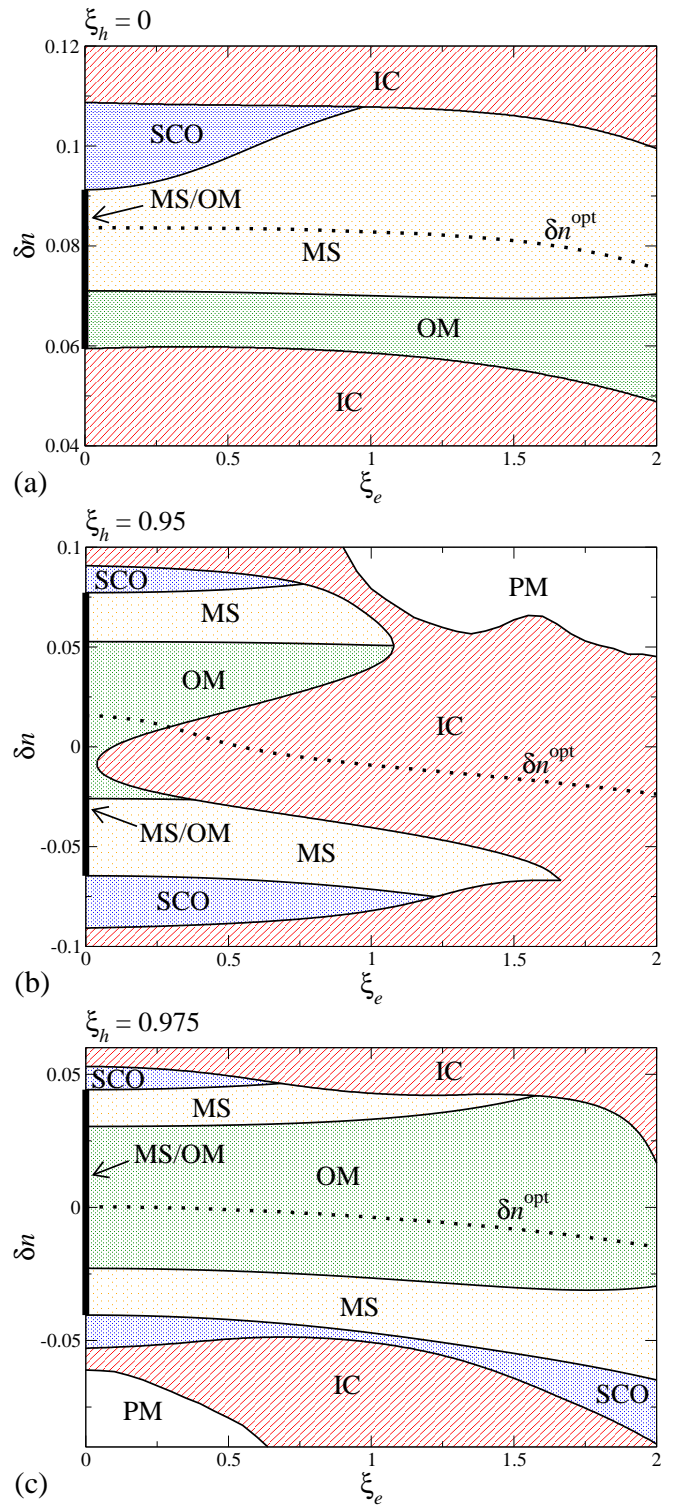

\begin{center}
\includegraphics[clip,width=\columnwidth]{pdTc.eps}\\
\includegraphics[clip,width=\columnwidth]{pdTc2.eps}\\
\includegraphics[clip,width=\columnwidth]{pdTc3.eps}
\end{center}
\caption{\label{pd}(color online) Magnetic order at $T=T_{N}^-$ as a
  function of $\xi_e$ and $\delta{n}$ for (a)
  $\xi_h=0$, (b) $\xi_h=0.95$, and (c) $\xi_{h}=0.975$.  Magnetic phases are as
  defined in the text. At $\xi_e=0$ the thick solid line indicates
  degenerate MS and OM solutions. For $\xi_e\neq0$, solid lines indicate
  phase boundaries, while the dotted line indicates the optimal doping
  $\delta{n}^{\text{opt}}$. }
\end{figure}

\begin{figure*}
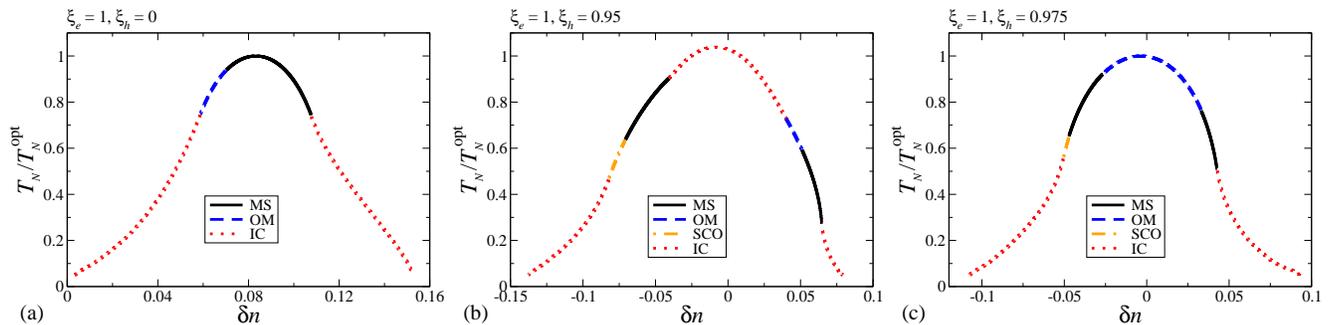

\begin{center}
\includegraphics[clip,height=4.2cm]{Tccuth0e1.eps}\hspace{0.1cm}\includegraphics[clip,height=4.2cm]{Tccuth095e1.eps}\hspace{0.1cm}\includegraphics[clip,height=4.2cm]{Tccuth0975e1.eps}
\end{center}
\caption{\label{tccut}(color online) Critical temperature $T_N$ of the SDW
  states as 
a function of doping $\delta{n}$ for $\xi_e=1$ and (a) $\xi_h=0$, (b)
 $\xi_h=0.95$, and (c) $\xi_h=0.975$. The temperature
is scaled by the maximum temperature for commensurate SDW order
$k_{B}T^{\text{opt}}_{N}=0.0646\, t_c$.}
\end{figure*}

The free energy in~\eq{eq:freeE} allows us to determine the phase diagram of the
model close to $T_{N}$. In~\fig{pd} we present phase diagrams
showing the ordered state realized at a temperature $T=T_{N}^{-}$
infinitesimally below $T_N$ as a function of $\xi_e$ and of the
doping relative to half-filling, $\delta{n}$, for various values of
$\xi_{h}$.   
In constructing the phase diagrams, we adjust $g_{\text{SDW}}$ such that for
each value of $\xi_{e}$ the maximum critical temperature of the
commensurate SDW as a function of the doping $\delta{n}$ is
$k_{B}T^{\text{opt}}_{N}=0.0646\,t_c$, which gives a reasonable ratio of $k_{B}T^{\text{opt}}_{N}$ to the
bandwidth. The variation of the optimal doping level $\delta{n}^{\text{opt}}$,
where $T_N$ is maximal,
with $\xi_e$ is shown by black dotted lines. The boundaries between the
different commensurate SDW phases are determined by the conditions on the
$\beta_0$, $\beta_{1}$ and $\widetilde{\beta}_{2}$ mentioned
above, where the coefficients in~\eq{eq:coefficients} were evaluated for
$g_{1}=g_2$ and using a
$1000\times1000$ ${\bf k}$-point mesh. In all phase diagrams we find
regions where IC SDW order occurs. Since the IC SDW ordering vector is
likely close to the commensurate SDW vector,~\cite{Schmiedt2011} the boundary
between the two phases is
determined by solving $1 + (g_{\text{SDW}}/V)\sum_{\bf
  k}(1/\beta)\sum_{n}G^{c (0)}({\bf k},i\omega_n)G^{f (0)}({\bf
  k}+{\bf Q}_1+{\bf \delta{q}},i\omega_n)=0$ for the critical temperature of
the ${\bf Q}_1+\delta{\bf q}$ SDW state, where $\delta{\bf q}=(0.01\pi/a,0)$,
$(0,0.01\pi/a)$. When the critical temperature of the ${\bf Q}_1+\delta{\bf
  q}$ SDW exceeds that of the commensurate SDW, an IC SDW is assumed to be
realized. We similarly find the critical doping for which there is no IC
SDW order, and the system remains paramagnetic (PM) down to zero
temperature. Note that we disregard states with $T_N<0.05\,T^{\text{opt}}_N$.

We first consider the phase diagram for $\xi_{h}=0$ [\fig{pd}(a)], which
corresponds to a Fermi surface with a single hole pocket and two electron
pockets at optimal 
doping as shown in~\fig{bs}(b).  A commensurate SDW 
state is realized here for $\delta{n}\approx\delta{n}^{\text{opt}}\pm0.025$
for all 
$\xi_e$. At $\xi_e=0$ the condition $\epsilon^{f}_{\bf k}=\epsilon^{f}_{{\bf
    k}+{\bf Q}_3}$ is satisfied, and hence the OM and MS states are
degenerate. These states have the lowest free energy at underdoping and near
optimal doping, but at overdoping the SCO is realized. Upon switching on a
finite $\xi_e$, the degeneracy of the MS and OM 
states is lifted, and the MS state is found to have the lower free
energy near optimal doping and at overdoping, while the OM state is
stable at underdoping. The SCO state is rapidly suppressed by a finite
$\xi_e$.

The phase diagrams for the case of two hole and two electron Fermi pockets
[see~\fig{bs}(c)] is
shown in Figs.\ \ref{pd}(b) and (c) for $\xi_h=0.95$ and $\xi_h=0.975$,
which correspond to hole pockets of lesser and greater
similarity, respectively. We note that the interaction strength
$g_{\text{SDW}}$ needed to produce the SDW state is roughly a third smaller
than for the single-hole-pocket scenario. 
For small $\xi_e$, a
commensurate SDW is nevertheless realized over a much greater doping range
than in the single-hole-pocket case. The OM phase is stable near optimal
doping, with 
the MS phase found at moderate doping, and the SCO found at
stronger doping. At larger values of $\xi_e$, however, we find a strong
tendency towards IC order in the $\xi_h=0.95$ case, with commensurate order
completely absent for $\xi_e>1.75$. In contrast, the commensurate SDW in
the $\xi_h=0.975$ case is present for all $\xi_e$ and is always realized about 
optimal doping.

In~\fig{tccut} we plot the critical temperature of the SDW states as a
function of doping $\delta{n}$ for constant-$\xi_e$ cuts through the three
phase diagrams in~\fig{pd}.
For the $\xi_h=0$ case [\fig{tccut}(a)] we note that there are substantial
IC SDW ``shoulders'' to the commensurate SDW dome, which extend up to
$T\approx0.75\,T^{\text{opt}}_N$. Although IC SDW states are also 
found at strong underdoping or overdoping in the 
$\xi_h=0.975$ case [\fig{tccut}(c)], they are realized over a
smaller doping range relative to the commensurate states and do not extend to
such high temperatures compared to the single-hole-pocket scenario. As shown
in~\fig{tccut}(b), however, slightly reducing $\xi_h$ leads to IC states
appearing at optimal doping.

\section{Discussion}

To summarize our main results, we have shown that in a two-band model of the
pnictides there are three distinct commensurate SDW states: The MS, OM, and
SCO phases. In a model with a single hole pocket and two electron pockets, the
MS state dominates the phase diagram, but the OM and SCO phases are possible
away from optimal doping. For a model with two hole pockets, in contrast, the
OM phase is stable at optimal doping, although the MS phase remains at under-
and overdoping. Since only the MS state is observed experimentally, we hence
conclude 
that the model with a single hole pocket gives a more reasonable description
of the physics. We nevertheless note that such a model displays a
rather strong tendency towards IC SDW order, which is not observed
experimentally.~\cite{Pratt2011} 

\begin{figure}
\begin{center}
\includegraphics[clip,width=\columnwidth]{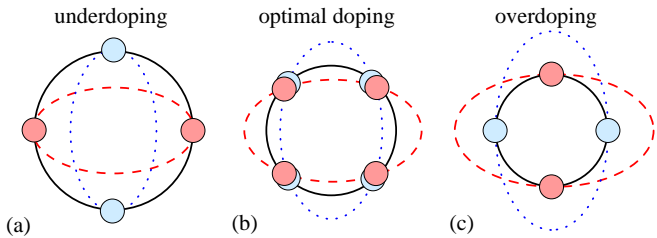}
\end{center}
\caption{\label{nesting}(color online) Schematic diagram of the nesting of the
two electron pockets (blue dotted and red dashed lines) with a single hole
pocket (black solid line). We show the situation for (a) underdoping, (b)
optimal doping, and (c) overdoping. The small shaded circles indicate the
region of best nesting of the electron pockets with the hole pocket. }  
\end{figure}

We consider the results for the single-hole-pocket case in more detail. In
agreement
with~\Ref{EC2010} we found that MS order was realized near optimal 
doping for arbitrarily small ellipticity of the electron Fermi
pockets. Away from optimal doping, however, states 
consisting of a 
superposition of commensurate SDWs with orthogonal ordering vectors
${\bf Q}_1$, ${\bf Q}_2$ are realized.
This can be understood via the following argument. At strong
underdoping, we expect that the best nesting between the
hole pocket and the elliptical electron pockets occurs for the states near the
major axis of the
electron pockets, as shown in~\fig{nesting}(a). Similarly, for strong
overdoping the best nesting occurs for the states near the minor axis of the
electron pockets [\fig{nesting}(c)]. In both cases, the SDW gaps for the two 
nesting vectors involve states far apart on the hole Fermi surface, and
so there should be little competition between them. Near optimal doping,
however, the nested electron Fermi pockets compete for the same states on the
hole Fermi surface [\fig{nesting}(b)], and hence it is more favorable for
only a single electron pocket to participate in the SDW. We note that the
variation of the nesting ``hotspots'' with doping is expected to have
significant consequences for the $a$-$b$ resistivity anomaly in the
pnictides.~\cite{Fernandes2011} 

The addition of a second hole pocket at the M point strongly affects
the physics. The doping range of commensurate SDW states is significantly
expanded, and the OM state is realized near optimal doping with a MS phase
appearing upon doping. This is consistent with previous investigations of the
two-orbital model.~\cite{Lorenzana2008,Raghuvanishi2011} In contrast, our
results are inconsistent with the degenerate OM and
MS phases found in~\Ref{EC2010}. This is due
to the fact that in the model of~\Ref{EC2010}
the $\Gamma$ hole pocket is mapped exactly onto the M
pocket by translation of ${\bf Q}_3$, i.e., the degeneracy condition
$\epsilon^{c}_{{\bf k}} = \epsilon^{c}_{{\bf k}+{\bf Q}_3}$ is always satisfied.
Despite the apparent differences, the phase diagram can be
understood in a similar way to the single-hole-pocket case. At optimal doping
the nesting of the electron pockets is not optimal for either hole pocket, but
instead corresponds to overdoping for the smaller pocket and underdoping for
the larger pocket, thus allowing the OM phase. Indeed, this perfectly
describes the nesting properties of the two-orbital model at
half-filling.~\cite{Brydon2011} Upon doping the system, the
nesting with one of the hole pockets is optimized, while the nesting is the
other hole pocket becomes extremely poor. As such, only a single hole pocket
participates in the nesting, and so the MS state is stable. The relative size
of the two hole pockets is thus crucial: If the two pockets are too dissimilar
in size, there will be a tendency for the $T_N(\delta n)$ curves in
Fig.~\ref{tccut} to split into two separate domes with commensurate order near
their maxima. The strong tendency towards
an IC SDW state at optimal doping suggests
that the $\xi_h=0.95$ case is close to this limit.
For somewhat lower $\xi_h$, i.e., when the hole pocket at the M point
is much smaller than the pocket at the $\Gamma$ point, the pocket at the M
point only shows good nesting with the electron pockets at very large
hole doping. At realistic doping levels, the smaller hole pocket is essentially
irrelevant for the SDW formation and the single-hole-pocket model is applicable.

Finally, we consider the implications of our results for the hypothesized
nematic state in the pnictides. Indeed, a major motivation for our study is
the connection of this phase with the MS SDW
state.~\cite{Fang2008,Kim2011,Fernandes2010} This can be 
seen via the following naive argument: After integrating out the CDW, we write
the free energy~\eq{eq:freeE} in terms of the sublattice magnetizations ${\bf
  m}_a$ and ${\bf m}_b$,
\beqarray
F &=& F_{0} + \frac{1}{2}\alpha\left(|{\bf m}_a|^2 + |{\bf m}_b|^2\right)
\notag \\
&& {}+ \frac{1}{16}(2\beta_0 + \beta_{1})\left(|{\bf m}_a|^2 + |{\bf
  m}_b|^2\right)^2 \notag \\
&& {}+ \frac{1}{4}(2\beta_0 - \beta_{1})\left({\bf m}_a\cdot{\bf
  m}_b\right)^2 \notag \\
&& {}+ \frac{1}{16}\widetilde{\beta}_{2}\left(|{\bf m}_a|^2 - |{\bf
  m}_b|^2\right)^2\,, \label{eq:FreeEmag}
\eeqarray
and identify the Ising nematic degree of freedom as $\varphi={\bf
  m}_a\cdot{\bf m}_b$. In the mean-field theory presented in this paper,
$\varphi$ is only non-zero in the MS phase. The inclusion of sufficiently
strong magnetic fluctuations, however, allows the nematic order parameter to
be non-zero \emph{above} $T_N$,\cite{Fang2008,Kim2011,Fernandes2010} so long
as the coefficient of $\varphi^2$ in~\eq{eq:FreeEmag} is negative, i.e
$2\beta_0<\beta_{1}$.  
This is the same condition that ensures that the MS
phase has lower free energy than the OM phase, and hence implies that a
nematic transition occurs at $T\geq
T_{N}$ when the SDW state shows stripe
order. It is
therefore intriguing that we find that $2\beta_0-\beta_{1}$ changes sign as
a function of doping in all cases. This suggests a strong
doping dependence of the
nematic phase: In particular, for a scenario with a single hole
pocket we expect that the nematic phase at underdoping will be weaker
and occur closer to $T_N$ compared to overdoping, or may even be 
absent. Since this is apparently not observed experimentally, our
results might imply that the magnetoelastic
coupling plays a more direct role in the structural
transition.~\cite{magph}

\section{Summary}

In this paper we have presented a weak-coupling study of the magnetic order in
a two-band model of the iron pnictides. Using the
Dyson equation for the Green's function of an arbitrary commensurate SDW
state treated at the mean-field level, we have obtained an expansion of the free
energy valid close to the
critical temperature. We have shown that this allows three commensurate SDW
states: The experimentally relevant stripe MS phase, the flux OM phase, and
the SCO phase for which only one sublattice orders. The competition of these
phases with one another and with IC SDW states has been  studied as a function
of the doping and the variation of key features of the non-interacting
electronic structure.

In particular, we have examined systems containing two elliptical electron
pockets and either a single hole pocket at the $\Gamma$ point or hole pockets
at both the $\Gamma$ and the M points. In the former case we find that the
MS state is
stable at optimal doping, while a superposition of two orthogonal SDW
states is stable away from optimal doping. In the latter scenario, however,
the OM phase is realized near optimal doping, while the MS state is stable at
moderate doping and the SCO state appears at higher doping levels. The
doping dependence of the associated phase diagrams can be understood in terms
of the changing nesting properties of the Fermi surface. Our results indicate
that the single-hole-pocket model is in better agreement with experiments,
presumably because any hole pocket at the M point in the real
materials is small and poorly nested with the electron pockets. The
single-hole-pocket picture
also suggest that the proposed nematic state in the pnictides should be
highly asymmetric with respect to electron vs.\ hole doping.

\acknowledgments

The authors thank A. Cano, M. Daghofer, I. Eremin, and I. Paul for useful
discussions. This work was financially supported in part by the Deutsche
Forschungsgemeinschaft through Priority Programme 1458.

\end{document}